\documentclass[12pt,a4paper]{revtex4}
\usepackage{amsmath}  
\usepackage{amssymb}  
\usepackage{dsfont}   

\begin{document}
\title{A quantum arrow of time}
\author{Alberto C. de la Torre}\email{delatorre@mdp.edu.ar}
\affiliation{
 Universidad Nacional de Mar del Plata\\
Mar del Plata, Argentina.}
\begin{abstract}
It is shown that position-momentum correlation is never decreasing and therefore it is a
good candidate as a quantum arrow of time devoid of shortcomings of other proposals.
 \\ Keywords: position momentum correlation, arrow of time.
\\ PACS: 03.65.Ca 03.65.Ta
\end{abstract}
\maketitle
\section{INTRODUCTION}
Since most fundamental laws of physics are time reversal invariant we need to refer to
some basic irreversible processes in order to fix a direction of time. So, several arrows
of time have been identified. The most widely used is the thermodynamic arrow based on
the never decreasing entropy of isolated systems. Other arrows of time related with
physical precesses are the cosmological arrow, the radiation arrow, the quantum arrow,
and the weak arrow.\cite{arrows} Interesting enough, the most obvious arrow of time, the
sense of becoming, is perhaps the less well understood. Whether all theses arrows are
equivalent and lead to a unique concept of time is an open question.

The proposed quantum mechanical arrow of time is based on the irreversible state collapse
occurring in a measurement. However this is not very convenient because the measurement
in a quantum system is related with an observer and is one of the most controversial
aspects in the foundations of quantum mechanics. In a better proposal, the quantum arrow
of time can be related with the decoherence of an initially coherent entangled state.
This has the inconvenience that it requires a particular class of states. In this short
note an alternative quantum arrow of time well understood, observer independent and state
independent is proposed, based on the never decreasing position-momentum correlation of a
free particle system.

\section{POSITION-MOMENTUM CORRELATION OBSERVABLE}
Given the position observable of a free particle $X$, the momentum observable $P$ is
defined by the operator that generates space translations. From this definition the
commutation relation $[X,P]=i\hbar$ can be derived. The position-momentum correlation is
defined as
\begin{equation}\label{Corr}
    C=\frac{1}{2}(XP+PX)\
\end{equation}
with commutation relations
\begin{equation}\label{ComRel}
   [X,C]= i\hbar X\ \hbox{ and }\ [P,C]= -i\hbar P \ .
\end{equation}

Position-momentum correlations have a simple explanation in an interpretation of quantum
mechanics suggested by quantum field theory. In this interpretation we can view the
``probability cloud'' as a permanent creation, propagation and annihilation of virtual
particles in an indefinite number making up the quantum field associated to some particle
type. We can think that the virtual particles are the components of the field that have
objective but ephemeral existence with position and momentum. In this view, Feynman
graphs are not only mathematical terms of a perturbation expansion but represent real
excitations of the quantum field.

Let us imagine then virtual components of the field created at a location at ``the
right'' of the one dimensional distribution $\rho(x) $, that is, with a \emph{positive}
value for the observable $X-\langle X\rangle$. If these components are moving with
momentum smaller than the mean value, that is, with \emph{negative} value for $P-\langle
P\rangle$ the relative motion will be towards the center and the distribution will
shrink. Similarly, the components created at the left and moving to the right have the
two offsets $X-\langle X\rangle$ and $P-\langle P\rangle$ with different sign, that is,
their (symmetrized) product is negative.

For simplicity, let us assume that in this state we have $ \langle X\rangle=\langle
P\rangle =0$ (the general state is obtained with the translation and impulsion operator).
Therefore the product of the two offsets in position and momentum is precisely the
correlation observable and the previous argument means that if the correlation is
negative the space distribution shrinks. We can prove this with rigour: let us calculate
the time derivative of the width of the distribution $\Delta^{2} x = \langle X^{2}\rangle
$. In the Heisenberg picture, assuming a nonrelativistic hamiltonian $H=P^{2}/2m$, we
have
\begin{equation}\label{shrink}
 \frac{dX^{2}}{dt}=\frac{-i}{\hbar}[X^{2},H]=\frac{-i}{2\hbar m}[X^{2},P^{2}]
 =\frac{1}{m}(XP+PX)=\frac{2}{ m}C.
\end{equation}
Taking expectation values we conclude that states with negative correlation shrink and
states with positive correlation expand, as expected from the heuristic argument given
above.

The momentum distribution for a free particle is time independent and if the state is
shrinking, that is, with negative correlation, we are approaching the limit imposed by
Heisenberg indeterminacy principle. This principle will not be violated because the
correlation will not remain always negative: at some time it will become positive and the
state will begin to expand. In fact, we can prove that the correlation is never
decreasing in time:
\begin{equation}\label{corrtimeincr}
 \frac{dC}{dt}=\frac{-i}{\hbar}[C,H]=\frac{-i}{4\hbar m}[XP+PX,P^{2}]
 =\frac{1}{m}P^{2}=2H,
\end{equation}
and this is a nonnegative operator. If a state is shrinking, at some later time it will
be spreading. Gaussian states of this sort have been reported\cite{rob} in a very
comprehensive paper.

Position-momentum correlation, like entropy in thermodynamics, is never decreasing and
can be used to define a quantum mechanical arrow of time without recourse any particular
class of states or to the controversial state collapse.

\end{document}